\documentclass[9pt,twocolumn,twoside]{pnas-new}

\templatetype{pnasresearcharticle} 
\usepackage{graphicx}
\usepackage{dcolumn}
\usepackage{bm}
\usepackage{float}
\usepackage{xcolor}
\usepackage{amsfonts}
\usepackage{multicol}

\title{Understanding, predicting, and tuning the fragility of vitrimeric polymers}

\author[a,1]{Simone Ciarella}
\author[a]{Rutger A. Biezemans}
\author[a,2]{Liesbeth M. C. Janssen}

\affil[a]{Theory of Polymers and Soft Matter, Department of Applied Physics,
Eindhoven University of Technology, Den Dolech 2, 5600 MB Eindhoven, The
Netherlands}

\leadauthor{Ciarella}

\significancestatement{
The intricate interplay between the molecular structure of a material and
its emergent macroscopic dynamics lies at the heart of modern materials
science. This interplay, however, remains notoriously poorly understood for
glass-forming systems.  The recent advent of vitrimers---a promising new class of
recyclable high-performance polymers with anomalous glass-forming properties
---revives the urgency to gain a full understanding of the elusive
structure-dynamics link in glassy matter. We combine computer simulations and
first-principles theory to rationally expose the structural origins of the
complex glassy dynamics in vitrimers. Our findings provide cues for a better
fundamental understanding of glass formation, and may stimulate the development
of novel, sustainable amorphous materials with on-demand functionalities. 
}

\authorcontributions{Please provide details of author contributions here.}
\authordeclaration{Please declare any conflict of interest here.}
\correspondingauthor{E-mail: \textsuperscript{1}S.Ciarella@tue.nl  \,\,  \textsuperscript{2}L.M.C.Janssen@tue.nl}


\begin{abstract}
Fragility is an empirical property that describes how abruptly a
glass-forming material solidifies upon supercooling.
The degree of fragility carries important implications for the
functionality and processability of a material, as well as for our fundamental 
understanding of the glass transition. However, the microstructural properties 
underlying fragility still remain poorly understood. 
Here, we explain the microstructure-fragility link in vitrimeric networks, a
novel type of high-performance polymers with unique bond-swapping functionality and unusual
glass-forming behavior.
Our results are gained from coarse-grained computer simulations and
first-principles Mode-Coupling Theory (MCT) of star-polymer vitrimers.  We
first demonstrate that the vitrimer fragility can be tuned over an
unprecedentedly broad range, from fragile to strong and even superstrong
behavior, by decreasing the bulk density.  Remarkably, this entire
phenomenology can be reproduced by microscopic MCT, thus challenging the
conventional belief that existing first-principles theories cannot account for
non-fragile behaviors.  Our MCT analysis allows us to rationally identify the
microstructural origin of the fragile-to-superstrong crossover, which is
rooted in the sensitivity of the static structure
factor to temperature variations. On the molecular scale, this behavior stems from a change in dominant
length scales, switching from repulsive excluded-volume interactions to
intrachain attractions as the vitrimer density decreases. Finally, we develop a
simplified schematic MCT model which corroborates our microscopically-founded conclusions and
which unites our findings with earlier MCT studies. Our work sheds new light
on the elusive structure-fragility link in glass-forming matter, and provides a
first-principles-based platform for designing novel amorphous materials with an
on-demand dynamic response.
\end{abstract}

\dates{This manuscript was compiled on \today}
\doi{\url{www.pnas.org/cgi/doi/10.1073/pnas.XXXXXXXXXX}}

\begin{document}

\maketitle
\thispagestyle{firststyle}
\ifthenelse{\boolean{shortarticle}}{\ifthenelse{\boolean{singlecolumn}}{\abscontentformatted}{\abscontent}}{}

\dropcap{V}itrimers are a new class of polymer glasses with exceptional
material properties, combining the malleability and recyclability of
thermoplastics with the high mechanical performance of
thermosets~\cite{montarnal2011silica, capelot2012metal}. Their unique processability stems from a
reversible bond-exchange mechanism, which allows covalent crosslinks in the
polymer network to change dynamically whilst preserving the total number of
bonds. In effect, vitrimers can thus behave as viscoelastic liquids at high
temperatures, and as crosslinked thermosets at low temperatures. Importantly,
the relaxation dynamics of vitrimers is characterized by two distinct
transition temperatures: a conventional glass transition temperature $T_g$ at
which the material undergoes kinetic arrest, and a topology-freezing transition
temperature $T_v$ at which the bond-switching rate becomes immeasurably slow.
The latter transition temperature can be controlled independently from $T_g$,
e.g.\ by tuning the concentration of catalyst for the exchange
reaction~\cite{denissen2016vitrimers}.

A widely employed classification scheme for glass-forming matter, 
proposed by Angell~\cite{Angell1995}, is based on the fragility---a measure for how rapidly the
relaxation dynamics of a material slows down upon approaching the glass
transition. The fragility index $m$ can be defined as $m=d \log_{10} \tau
/d(T_g/T)|_{T=T_g}$, where $\tau$ is the structural relaxation time and $T$ is
the temperature. For so-called 'strong' glass formers, the relaxation-time
growth near $T_g$ follows an Arrhenius law, while for 'fragile' glass formers
the relaxation time increases more abruptly upon cooling (super-Arrhenius
behavior), giving rise to larger values of $m$.  These differences in the
fragility index are important for two reasons.  First, the concept of fragility
offers a unifying framework to classify seemingly disparate classes of
glass-forming matter, ranging from atomic liquids~\cite{Angell1995} to
colloidal~\cite{Mattsson2009} and cellular systems~\cite{Sussman2018}. As such,
it is paramount to an ultimately universal description of the glass transition.
Second, the fragility index of a material directly impacts its functionality and
processability: a low $m$, for example, generally implies a broader glass
transition temperature range, increasing the ease with which a material can be
molded. Remarkably, while most polymers are fragile \cite{SokolovJCP2016},
vitrimers exhibit an even lower fragility index than the prototypical strong
glass former silica~\cite{capelot2012metal}, rendering them 'superstrong' and highly suitable for
(re)processing. Despite both the fundamental and practical importance of
fragility, however, there is still no detailed understanding of the physical
mechanisms underlying this phenomenon, and a predictive theory that can
rationally link the observed fragility to a material's microstructure is still
lacking~\cite{debenedetti2001supercooled,Dyre2006,Cavagna2009,Tarjus2011,Berthier2011,binder2011glassy,Charbonneau2017}.
This constitutes not only a major fundamental problem, but it also
impedes any rational approach toward the development of sustainable, high-performance
amorphous materials with an on-demand dynamic response.

Empirical studies over the last few decades have nonetheless yielded valuable
insight into the molecular and microstructural origins of fragility. It is now
well established that network-forming materials such as silica, which are
characterized by anisotropic interactions, tend to behave as strong glass
formers~\cite{saika2001fragile,Sastry1998}, while fragile materials are often
governed by isotropic and excluded-volume interactions~\cite{Krausser13762,Ozawa_2016}.  In dense colloidal
suspensions, the fragility is also intimately related to the particle softness~\cite{GnanNature},
roughness \cite{Asai2018}, elasticity
\cite{Yan6307,VanDerScheer2017,Mattsson2009,Philippe2017}, and confinement effects~\cite{doi:10.1063/1.4905472}.  
For polymeric
systems, the glassy dynamics is generally more complex, owing to the
competition between intra- and interchain degrees of
freedom~\cite{chong2007structural,Aichele2004}. A recent study
\cite{SokolovJCP2016} argues that in fragile polymers with high molecular
weights, the slow intrachain (segmental) relaxation dynamics can give rise to
unusually high values of $m$, even though the interchain relaxation may remain
comparable to that of moderately fragile liquids.  Finally, the liquid
fragility also correlates well with material properties deep in the glass
phase, in particular the vibrational properties and nonergodicity factor at
temperatures well below $T_g$ \cite{Scopigno2003}.

From the theoretical side, it is widely accepted that material-dependent
dynamical properties such as the fragility must ultimately be related to
structural ones~\cite{Royall2015}.
After decades of intense research, there now exist numerous
theories of the glass transition that aim to rationalize the observed fragility
for a given material composition and structure, notably including Random First
Order Transition Theory (RFOT) \cite{Xia1999}, geometric frustration
\cite{Tarjus2005a,Sausset2008}, and Mode-Coupling Theory (MCT)~\cite{gotze2008complex,Gotze1992}. Among these, MCT
stands out as the only theory that is founded entirely on first principles. As
such, MCT seeks to predict the full microscopic relaxation dynamics of a
glass-forming system (as a function of time, temperature, density, and
wavenumber $k$) in a semi-quantitative way~\cite{Kob2002,Gotze1999} based solely on knowledge of simple structural material
properties such as the static structure factor $S(k)$. 
However, even though experiments suggest that subtle features in $S(k)$, especially in the main peak
at wavenumber $k=k_0$, may correlate well with the fragility index
\cite{Mauro2014}, MCT often fails to give an accurate prediction of the
observed fragility. In fact, due to the inherent mathematical form of the 
MCT equations, the theory is only capable of predicting a fragile (power-law) divergence 
of $\tau$ near the MCT glass transition. 
Indeed, a 2011 review contemplated that 
a quantitative account of fragility from a purely first-principles
microscopic theory remains, thus far, only a ``dream'' \cite{Tarjus2011}.

Here we challenge this status quo by elucidating and predicting the
fragility of vitrimeric polymers using coarse-grained Molecular Dynamics (MD) simulations
and MCT. We first demonstrate that, within our vitrimeric model system, the
fragility index can be tuned over an unprecedentedly broad range,
from fragile super-Arrhenius to superstrong sub-Arrhenius
behavior, by tuning the bulk density. Furthermore, we find that the 
fragility does not change with the swap rate, suggesting that the observed phenomenology is more related 
to the intrinsic structure of the liquid rather than the swap-dependent dynamics.
Remarkably, we can reproduce
and predict this entire range of fragilities from microscopic MCT, 
using only the static structure factors at the
corresponding temperatures and densities as input.  Our first-principles MCT
analysis allows us to rationalize the observed crossover from fragile to
superstrong behavior in terms of a change in dominant length scales, which
switches from inter- to intrachain interactions as the density decreases.
Moreover, we can directly attribute the fragility at a given density to the
growth behavior of the static structure factor: at low
densities, anomalously strong vitrimers are governed by an anomalously weak
variation of $S(k)$ upon cooling. This new insight finally allows
us to develop a simple schematic (wavevector-independent) MCT model which not
only recovers our fully microscopic MCT predictions, but which also reveals how
our MCT-predicted fragility range can be united with MCT's well-established
fragile power-law divergence.  Overall, this work sheds new light on the
microstructural origins of fragility, and uncovers a previously unknown
application domain for a near-quantitative first-principles theory of glassy
dynamics. The here established rational link between molecular length
scales on the one hand, and the macroscopic material response on the other
hand, may also guide ongoing experimental efforts to design and optimize
new structural glasses with tunable dynamic functionalities.

\section{Approach}
Our results are gained from MD simulations, which serve as 'numerical
experiments', combined with theoretical predictions from both fully microscopic
and schematic MCT.  The simulation model consists of a binary mixture
of star polymers, inspired by the system used in Ref.\ \cite{Ciarella2018} (see
Fig.\ 1 and Materials and Methods). 
The star-shaped structure, which is comparable to soft colloids~\cite{Likos2006,PhysRevLett.82.5289,PhysRevLett.80.4450},
offers a complex scenario for fragility~\cite{Asai2018,Yan6307,VanDerScheer2017,Mattsson2009,Philippe2017,GnanNature}, 
on top of tunable mechanics~ \cite{Gu2017}.
Briefly, the monomer segments are connected through
harmonic springs with equilibrium length $L_0$; steric repulsion among non-nearest-neighbor segments 
is implemented through a Weeks-Chandler-Andersen (WCA) potential with cutoff radius $\sigma=0.9L_0$. 
Importantly, the monomers at the ends of the arms are
functionalized such that they can form attractive bonds which can swap reversibly,
thus endowing them with vitrimeric functionality \cite{Ciarella2018,sciortino2017three}. 
For these attractive end-segment interactions, we use a generalized 20-10 Lennard-Jones potential 
with minimum energy $\epsilon\gg k_\mathrm{B}T$, where $k_\mathrm{B}$ is the Boltzmann constant.
We assume that every bond-swapping reaction requires an activation energy $\Delta
E_{swap}$, which experimentally can be controlled by e.g.\ the amount of
catalyst. Physically, by varying $\Delta E_{swap}$, we can thus tune the
location of the topology-freezing transition temperature $T_v$ independently
from $T_g$. In the remainder of this paper, we adopt reduced units for the length, energy, and monomer mass  
such that $L_0=1$, $\epsilon=1$, and $m_0=1$, respectively; the temperature $T$ is defined in units of $\epsilon/k_\mathrm{B}$.
 
To quantify the time-dependent glassy relaxation dynamics, we measure the
monomer-averaged self- and collective intermediate scattering
functions~\cite{Aichele2004} $F_s(k,t) = N^{-1} \langle \sum_{j=1}^N \exp\left[
-i\bm{k}\cdot \left( \bm{r}_j(0)-\bm{r}_j(t) \right) \right] \rangle$ and $F(k,t) = N^{-1} \langle \sum_{j,l=1}^N \exp\left[
-i\bm{k}\cdot \left( \bm{r}_j(0)-\bm{r}_l(t) \right) \right] \rangle$, where
$N$ is the total number of monomers, $\bm{k}$ is a wavevector of magnitude $k$,
$\bm{r}_j(t)$ denotes the position of monomer $j$ at time $t$, and the brackets
$\langle \hdots \rangle$ represent an ensemble average. Note that this
monomer-averaged approximation, which has been successfully applied to
coarse-grained glassy
polymers~\cite{chong2007structural,Aichele2004,chong2002mode,Sharpe1999,schweizer1997integral},
would correspond in experiment to the case where all segments have identical
scattering form factors.  The vitrimer microstructure is quantified by the
average static structure factor $S(k) = N^{-1} \langle \sum_{j,l=1}^N
\exp\left[ -i\bm{k}\cdot \left( \bm{r}_j(0)-\bm{r}_l(0) \right) \right]
\rangle$.  We define the structural relaxation time $\tau$ as
$F_s(k_0,\tau)=e^{-1}$ and the operational glass transition
temperature $T_g$ as the temperature below which $F_s(k,t)$ fails to decay
within the maximum simulation time of $10^{4.5}\tau_0$, i.e.\ $\tau(T_g)=10^{4.5}\tau_0$. 
Here $\tau_0$ is the structural relaxation time of a reference
fluid state at density $\rho=1.58$ and temperature
$T=0.2$, which corresponds to $\tau_0=0.06 [(m_0 L_0^2/\epsilon)^{1/2}]$.  
The fragility index $m$
can subsequently be extracted from $\tau(T_g)$ for any given density.  To
investigate the role of the topology-freezing transition temperature $T_v$, we
perform simulations both for $\Delta E_{swap}=0$, implying barrierless bond
swaps, and $\Delta E_{swap}=\infty$, implying $T_v \gg T \ge T_g$.

In order to elucidate the link between the vitrimer microstructure and its
time-dependent relaxation dynamics, we invoke fully microscopic ($k$-dependent)
Mode-Coupling Theory (see Materials and Methods). Briefly, MCT amounts to
a first-principles-based non-linear equation for the dynamic intermediate scattering
function, which can be solved self-consistently once the static structure 
factor $S(k)$ for a given temperature and density is known \cite{gotze2008complex,Liesbeth2018front}.
Though it is well known that MCT is not exact~\cite{Charbonneau2011}, and is generally
only quantitatively accurate at temperatures well above $T_g$, 
the theory allows us to identify---at least
qualitatively---which microstructural features dominate the glassy dynamics
manifested in $F_s(k,t)$ and $F(k,t)$.
As detailed below, our theoretical analysis reveals a direct 
correlation between the broad fragility range of vitrimers and the particular sensitivity of $S(k)$
to temperature variations, effectively captured by the anomalous growth of the structure factor peak $S(k=k_0)$.
To unambiguously confirm this 
unique structure-fragility link, we also develop a minimal MCT model that
takes into account only the growth of $S(k_0)$; by showing that the model can 
reproduce all qualitative features of the fully microscopic MCT predictions,
we can thus rationally establish how different apparent fragilities 
emerge from different microstructures.

\begin{figure}
\includegraphics[width=0.49\textwidth]{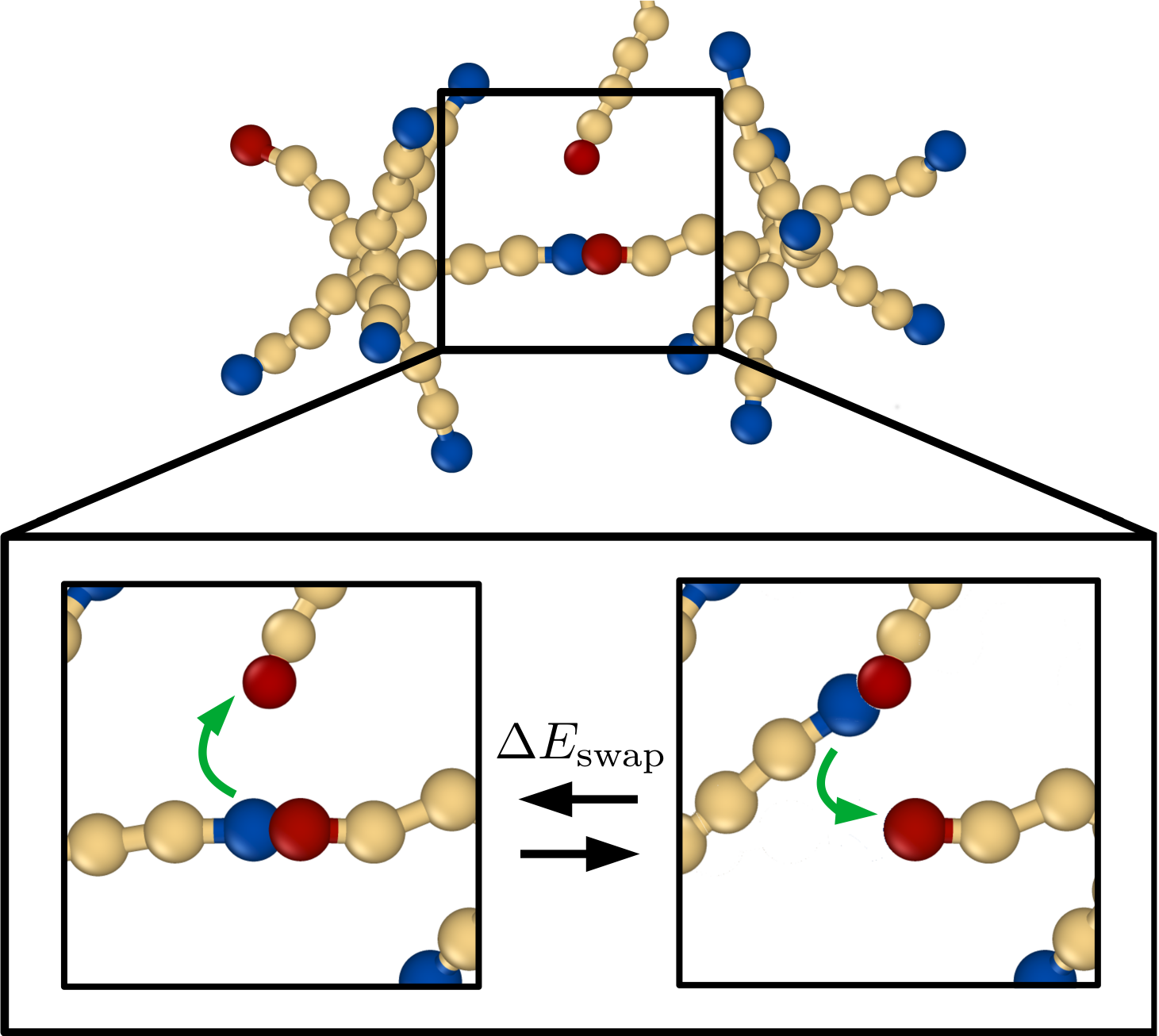}
\caption{Illustration of the vitrimer simulation model. 
The system is composed of a mixture of 8-arm star polymers, the monomer ends of which (shown as blue and red beads)
are capable of forming swappable vitrimeric bonds. The activation energy barrier
for a bond-swap event is $\Delta E_{\text{swap}}$; 
in our simulations, we can tune the value of $\Delta E_{\text{swap}}$ between $0$ and $\infty$.}
\label{f_model}
\end{figure}

\section{Results}

\subsection{Dynamics from numerical simulation}

\begin{SCfigure*}
\includegraphics[width=0.7\textwidth]{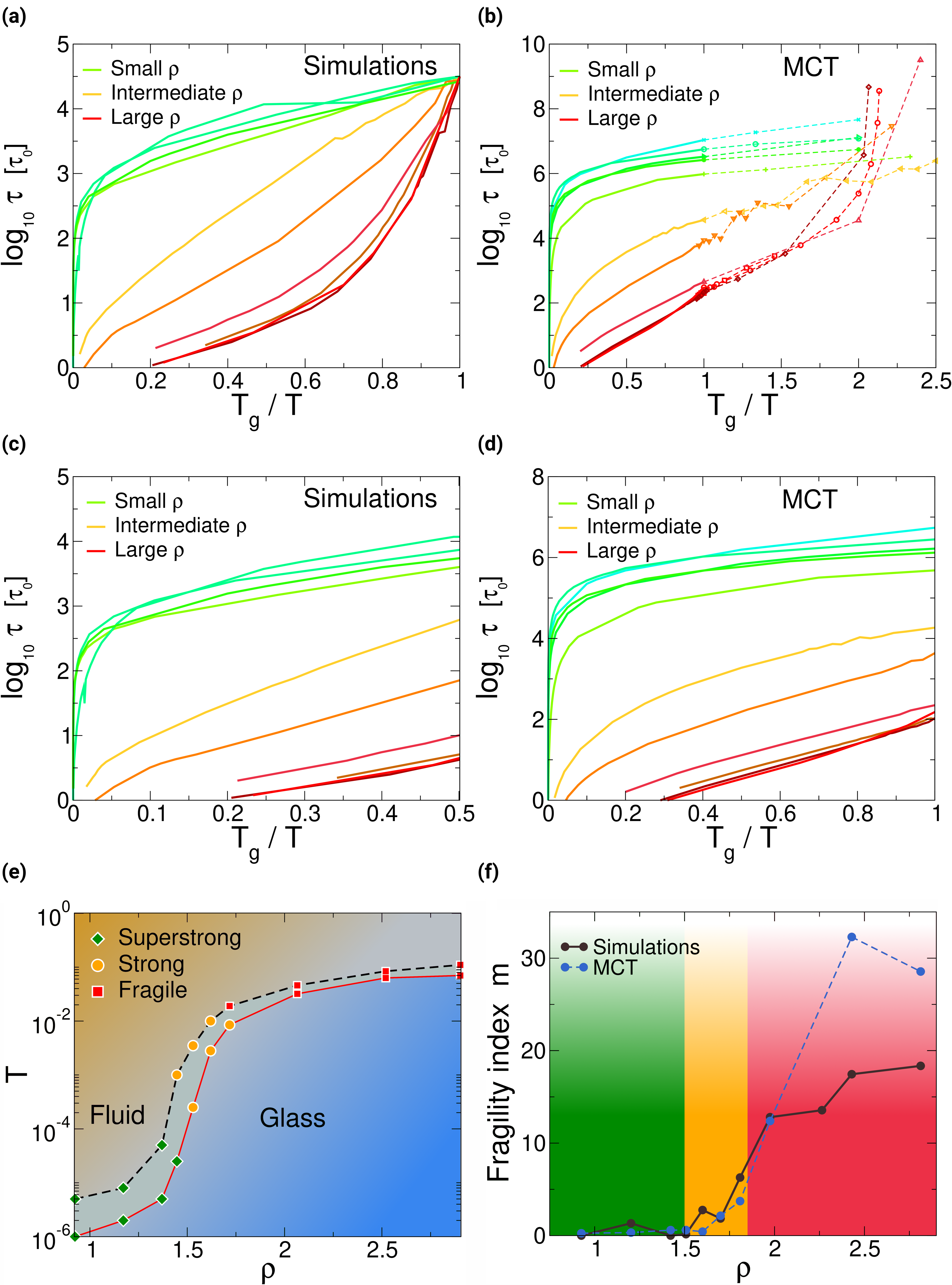}
\caption{(a,b) Angell plot of the structural relaxation time $\tau$ versus normalized inverse temperature, evaluated at the peak $k_0$. 
Data are obtained from (a) MD simulations and (b) fit-parameter-free microscopic MCT. Different curves correspond to different bulk densities.
The activation barrier for bond swapping was set to $\Delta E_\text{swap}=0$.
In both panels, the glass transition temperature $T_{g}$ is set as the temperature at which the simulated system falls out
of equilibrium such that $\tau = 10^{4.5} \tau_0$.
(c) Same MD data as in (a), with the temperature axis limited to $0.5T_g$. 
(d) Same MCT data as in (b), with the temperature axis limited to $T_g$. We estimate
that $T_{g,MCT}\approx 0.5T_g$ for the most fragile samples. 
(e) Simulated phase diagram as a function of temperature $T$ and bulk density $\rho$.
The phase labeled as 'fluid' corresponds to an underconstrained network of stars in which the dynamics is still ergodic such that
$\tau < 10^{4.5} \tau_0$. The phase labeled as 'glass' corresponds to the non-ergodic phase with $\tau > 10^{4.5} \tau_0$.
The black dashed line corresponds to $\Delta E_{swap}=\infty$ and the red line to $\Delta E_{swap}=0$. 
The region shaded in grey indicates the regime in which the glass can be melted by vitrimeric
bond swaps.
(f) Fragility indices $m$ as a function of bulk density $\rho$, obtained from polynomial fitting of 
the simulated and MCT-predicted Angell plots (panels a and b) near the respective glass transition temperatures.
The regions shaded in green, orange,
and red correspond to superstrong, strong, and fragile behavior, respectively.}
\label{fig_4panels}
\end{SCfigure*}

We first discuss the vitrimer relaxation dynamics obtained from simulation.
Figure~\ref{fig_4panels}(a) shows the Angell plot of the structural relaxation times
$\tau$ as a function of normalized inverse temperature, evaluated at the peak $k_0$.
It can be seen that high-density vitrimers ($\rho \geq 2.0$) 
exhibit a relatively steep growth of $\tau$ upon supercooling, indicative of fragile glass
formation, while intermediate densities ($1.5 < \rho < 2.0 $) yield an Arrhenius form that is
characteristic of strong network-forming materials. Notably, as the bulk
density decreases even further ($\rho< 1.5 $), the vitrimer melt approaches the glass
transition with remarkably weak temperature dependence---a sub-Arrhenius
behavior which we designate as superstrong. The respective glass transition
temperatures change from $T_{g} \approx 10^{-1}$ 
at high densities to $T_{g}\approx10^{-6}$ at the lowest
densities studied. In Fig.\ S2 we show the same plot for different values of $k$, confirming
that the fragility is $k$-independent.

To quantify the observed fragility, we also plot the fragility
indices $m$ as a function of bulk density in
Fig.~\ref{fig_4panels}(f); the $m$ values are derived from polynomial fitting of the data
in Fig.~\ref{fig_4panels}(a). 
The superstrong functionality at low densities is manifested in the
anomalously low values of $m$, revealing an essentially constant fragility of
$m \approx 1$ for $\rho_c \leq 1.5$. As the density increases beyond $\rho_c$, 
the fragility index increases monotonically and the material becomes
increasingly more fragile. Hence we may conclude that, for our vitrimer
simulation model, the bulk density provides a thermodynamic control parameter
to tune the fragility from superstrong to strong to fragile.

The influence of $T_v$ can be seen in the ($\rho, T$) phase diagram of Fig.~\ref{fig_4panels}(e).
Our simulations reveal two different amorphous phases: an ergodic phase
in which $F_s(k,t)$ fully decays to zero within the time scale of simulation
(marked as 'fluid' in Fig.~\ref{fig_4panels}(e)), and a solid glass phase in which $F_s(k,t)$
remains finite such that $\tau \geq 10^{4.5} \tau_0$. 
For all the ($\rho, T$)
state points considered here, the star polymers form a network kept together by
the functionalized attractive end segments. 
Our mixture composition ensures that
the maximum number of inter-star bonds (set by the minority end species, see Materials and Methods)
is below the rigidity transition~\cite{doi:10.1080/14786446408643668,CALLADINE1978161,Phillips1985}, 
so that the system can relax without undergoing bond swaps; this is confirmed by our 
$\Delta E_{swap}=\infty$ results.
Furthermore, this mixture choice allows us to explicitly decouple the glass transition $T_g$ from the 
topology-freezing transition temperature $T_v$: we can reach $T_g$ for both $T_g < T_v$ and $T_g > T_v$.
Note that in the earlier work of Ref.~\cite{Ciarella2018}, the relaxation time would diverge at 
$T_v$~\cite{denissen2016vitrimers} such that $T_g = T_v$.
Figure \ref{fig_4panels}(e) shows
the ergodic-to-glass transition line (i.e., the density-dependent values of
$T_g$) for the two extreme cases of $\Delta E_{swap}=0$ and $\Delta
E_{swap}=\infty$, implying topology-freezing transition temperatures of $T_v
\rightarrow 0$ and $T_v \rightarrow \infty$, respectively. It can be seen that,
while the shape of the two transition lines is similar, the introduction of
reversible bond swaps ($\Delta E_{swap}=0$) shifts the glass transition to
lower temperatures and higher densities.  Thus, for a given $T_g$, 
decreasing $T_v$ leads to a higher critical density at which vitrification
occurs, and therefore to a melting of the glass within the gray-shaded region of Fig.~\ref{fig_4panels}(e).
This effect is not surprising, as enhanced vitrimeric functionality is known to
lead to more efficient relaxation and more liquid-like behavior
\cite{Ciarella2018,rovigatti2018self}. 
Interestingly, we find that the fragility index $m$ along the
glass transition line is virtually unaffected by our choice of $\Delta
E_{swap}$ (Fig.\ S4), implying that the fragility is determined by intrinsic star polymer
properties, rather than by bond-swapping events. Since the quantitative
difference between $T_v \rightarrow 0$ and $T_v \rightarrow \infty$ in the
phase diagram is relatively minor, we focus in the remainder of the paper only
on the data for $\Delta E_{swap}=0$, for which we have better statistics.

\subsection{Dynamics from Mode-Coupling Theory}

Let us now investigate how certain properties of the microstructure may underlie and 
explain the unusually broad fragility range observed in simulation.  To this
end, we perform a theoretical analysis based on MCT, in which we use
the simulated static structure factors $S(k)$ at a given $(\rho,T)$ state point 
as input to predict the full dynamics. 
While decades of research
have firmly established that MCT can only predict a fragile power-law
divergence of $\tau$ near the MCT critical point \cite{gotze2008complex,Leutheusser1984}, the microstructural
information encoded in $S(k)$ may be sufficient to account at least
qualitatively for the observed differences in fragility indices
\cite{Landes2019,Coslovich2012}. 
The results of our microscopic, fit-parameter-free MCT calculations for the single-particle dynamics 
are shown in the Angell plot of Fig.~\ref{fig_4panels}(b).
Similar plots for the collective dynamics are presented in Fig.\ S3. 
We first point out that the respective $T_g$ values used in the Angell
plot are obtained from MD simulations; these temperatures need not coincide
with the MCT-predicted glass transition $T_{g,MCT}$. In fact, we find that
MCT predicts liquid-like behavior at all the state points considered in simulation, implying 
that the vitrimeric MCT transition temperature must lie \textit{below} the
simulated $T_g$. This finding is also consistent with other site-averaged MCT polymer 
studies~\cite{Frey2015,chong2007structural,Baschnagel_2005}. Based on extrapolation into
the glassy regime (dashed lines in Fig.~\ref{fig_4panels}(b)), we estimate that  
$T_{g,MCT}\approx 0.5 T_g$ for the most fragile samples.
In order to enable a fair comparison between the MCT and simulation data in the well-equilibrated regime,
we also present rescaled Angell plots in Figs.~\ref{fig_4panels}(c) and (d), corresponding to temperatures up to $0.5T_g$ 
in simulation and $\sim0.5T_{g,MCT}$ in MCT, respectively. 

The MCT results in Fig.\ \ref{fig_4panels} show that, despite the underestimation of the glass transition temperature, the theory
is remarkably successful at capturing all \textit{qualitative} features of the Angell plot. Indeed, in agreement with simulation,
MCT predicts fragile-like
 behavior at high vitrimer densities, and strong and even anomalously superstrong behavior at low bulk densities. This trend is already observed
in the high-temperature regime (Figs.~\ref{fig_4panels}(c,d)), in which all MCT-input structures are properly 
equilibrated. The MCT data in the glassy regime, obtained from fit-parameter-free MCT calculations 
using the simulated $T<T_g$ static structure factors as input (dashed lines in Fig.\ \ref{fig_4panels}(b)), indicate a continuation of the same fragility trend. However, it must be noted 
that this MCT extrapolation inherently suffers from quantitative inaccuracies, as the corresponding microstructures at $T<T_g$ are poorly equilibrated.
Moreover, our microscopic MCT analysis is based on 
monomer-averaged (structural and dynamical) correlation functions;
an explicit multi-component treatment that distinguishes between 
different monomer species would possibly 
yield a higher MCT glass transition temperature $T_{g,MCT}$ \cite{Frey2015,chong2007structural}.
Nonetheless, the fact that MCT manifestly predicts different growth behaviors of $\tau(T)$ for different bulk densities, consistent with
simulation, suggests that MCT can provide a meaningful framework to elucidate the non-trivial structure-fragility link in vitrimers.
To complete our comparison, we also present the MCT fragility indices $m$, extracted from the most glassy MCT data of Fig.\ \ref{fig_4panels}(b).
As can be seen in Fig.\ \ref{fig_4panels}(f), these MCT results are also in good qualitative agreement with numerical experiment across the entire
density range studied. Notably, MCT accurately reproduces the crossover
density $\rho_c\approx 1.5$ at which the vitrimer fragility changes from superstrong to
strong. 

At first glance, the fact that microscopic MCT can predict the full
fragility range of our vitrimer system may seem at odds with the fragile
power-law divergence expected from MCT on general mathematical grounds. However,
as we discuss below and confirm explicitly using a schematic MCT model, 
such a divergence appears only sufficiently close to the MCT glass transition temperature $T_{g,MCT}$.
For the vitrimer system under study, the MCT critical point falls outside the ($\rho,T$) domain 
accessible in simulation, rendering the theory capable of predicting \textit{manifestly} different 
fragilities. We now use these MCT results to establish---on a rational, first-principles
basis---which specific features of $S(k)$ underlie the observed fragility range of vitrimers.

\begin{SCfigure*}
\includegraphics[width=0.7\textwidth]{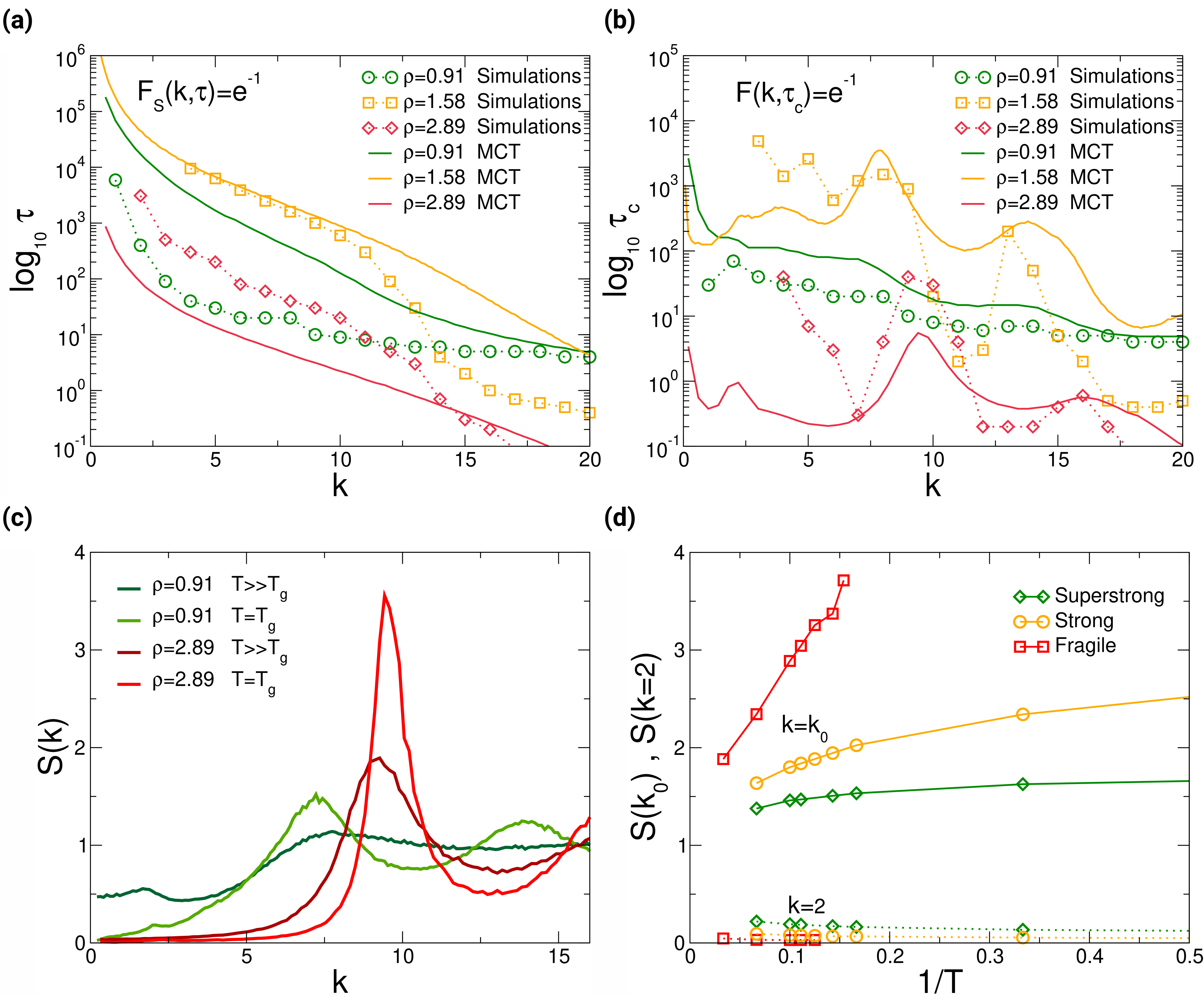}
\caption{(a) Averaged single-particle relaxation times $\tau$ as a function of wavenumber $k$ for different densities at temperatures close to $T_g$,
obtained from simulation (dotted lines) and full MCT (solid lines). 
(b) Collective relaxation times $\tau_c$ as a function of wavenumber $k$ for the same densities and temperatures as in panel (a), 
obtained from simulation (dotted lines) and full MCT (solid lines).
(c) Static structure factors for high-density ($\rho=2.89$) and low-density ($\rho=0.91$) vitrimers, measured both close to the glass transition temperature
and well above $T_g$. The main peak shifts towards smaller $k$ upon decreasing the bulk density of the system. 
(d) Temperature-dependent change in the value of $S(k)$ along different isochores, measured at both the main peak position $k_0$ (solid lines) and $k=2$ (dotted lines). 
By supercooling the vitrimer system at constant density, the peak growth at $k_0$ is 
most pronounced at high density (red line, $\rho=2.89$), and becomes weaker when the 
density decreases (orange line, $\rho=1.58$). At low density (green line, $\rho=0.91$), the temperature-dependent 
growth of the peak is remarkably weak. 
}
\label{fig3}
\end{SCfigure*}

\subsection{Linking structure to fragility}

The key structural quantity in our analysis is the monomer-averaged static
structure factor $S(k)$, which probes static correlations between two density
modes at wavenumber $k$. In simple dense glass-forming systems, the slowest (i.e.\ most glassy) density modes
correspond to the main peak of $S(k)$, i.e., $k_0\approx 2\pi/\sigma$,
where $\sigma$ denotes the typical monomer-particle diameter.
MCT provides a general interpretation of this phenomenon in terms of the cage effect
\cite{Charbonneau13939,charbonneau2014hopping}: as $S(k_0)$ grows upon
vitrification, particles become more confined within local cages formed by
their nearest neighbors. This local dynamic slowdown, in turn, drives the
slowdown at all other length scales via non-linear mode coupling, ultimately
resulting in complete kinetic arrest.
On the other hand, network-forming systems may be governed by larger length scales
that are related to, e.g., the intrinsic network topology or crosslinker distance~\cite{rovigatti2018self}. 
Hence, to elucidate the dominant microstructural
origin of the observed fragility of vitrimers, we must consider the structure and dynamics at both $k_0$ and smaller values of $k$.

Figure~\ref{fig3} shows the explicit $k$-dependence of both the structural relaxation times and the vitrimeric static structure factors
for several densities and temperatures in the well-equilibrated regime. Let us first consider the single-particle relaxation times $\tau(k)$ 
and the collective analogues $\tau_c(k)$ in Figs.\ \ref{fig3}(a) and (b), respectively, which are defined as $F_s(k,\tau)=e^{-1}$ and $F(k,\tau_c)=e^{-1}$. 
Interestingly, while the self-dynamics is the slowest at low $k$-values for all densities considered, the 
\textit{collective} dynamics in Fig.\ \ref{fig3}(b) reveals a more complex $k$-dependent picture~\cite{roldan2017connectivity}. 
We find that fragile high-density systems are 
governed by the typical caging mode at $k_0$, 
essentially conforming to simple glass-forming behavior (red lines in Fig.\ \ref{fig3}(b)), 
whereas superstrong low-density systems lack a notable dynamic signature at $k_0$ and are dominated by low-$k$ modes instead 
(green lines in Fig.\ \ref{fig3}(b)). 
This trend, i.e.\ the relative decrease of $\tau_c(k_0)$ with decreasing fragility, is observed both in simulation and in full MCT. 
Since the single-particle dynamics is enslaved to the collective relaxation dynamics, 
these qualitative differences in collective $k$-dependent relaxation indicate an important change in 
the underlying relaxation mechanism: as the bulk density decreases and 
the material becomes stronger, the role of the conventional cage effect associated with $k_0$ diminishes. The fact that MCT also 
qualitatively predicts this change on first-principles grounds clearly suggests that the key explanation for this 
unusual phenomenology must lie in $S(k)$.

Figure~\ref{fig3}(c) shows the full $S(k)$ curves for several representative vitrimer samples. 
We first note that the location of the $S(k_0)$ peak 
shifts toward lower wavenumbers and that the value of $S(k\rightarrow 0)$ increases as the density decreases.
The latter is also consistent with the fact that a low-density polymeric network should have a relatively high compressibility.
However, we find that the growth behavior of the $S(k_0)$ peak height upon supercooling is qualitatively different for low- and high-density vitrimers.
As can be seen in Fig.~\ref{fig3}(d), fragile vitrimers at high densities exhibit a relatively dramatic $S(k_0)$ temperature response and vitrify
over a narrow temperature range. This abrupt growth of $S(k_0)$ is similar to the fragile behavior of, e.g., compressed hard spheres; indeed, we have verified 
that isothermal compression in our vitrimer model also yields a fragile growth of $\tau$ upon approaching the
glass transition density, both in simulation and in MCT (see Fig.\ S5). 
Conversely, at low densities, the $S(k_0)$ height becomes remarkably insensitive to changes in the temperature,
growing only very weakly across the entire range of supercooling accessible in simulation. In fact, the temperature-dependent change 
in $S(k_0)$ for these superstrong samples is almost comparable in magnitude to the $S(k)$ change at much lower wavenumbers. Based on this analysis, 
we can thus conclude that the diminishing of the cage effect in low-density vitrimers, and the corresponding emergence
of superstrong fragilities, is rooted in the anomalously weak temperature-dependent variation of the static structure factor peak $S(k_0)$.

\begin{figure}
	\includegraphics[width=0.49\textwidth]{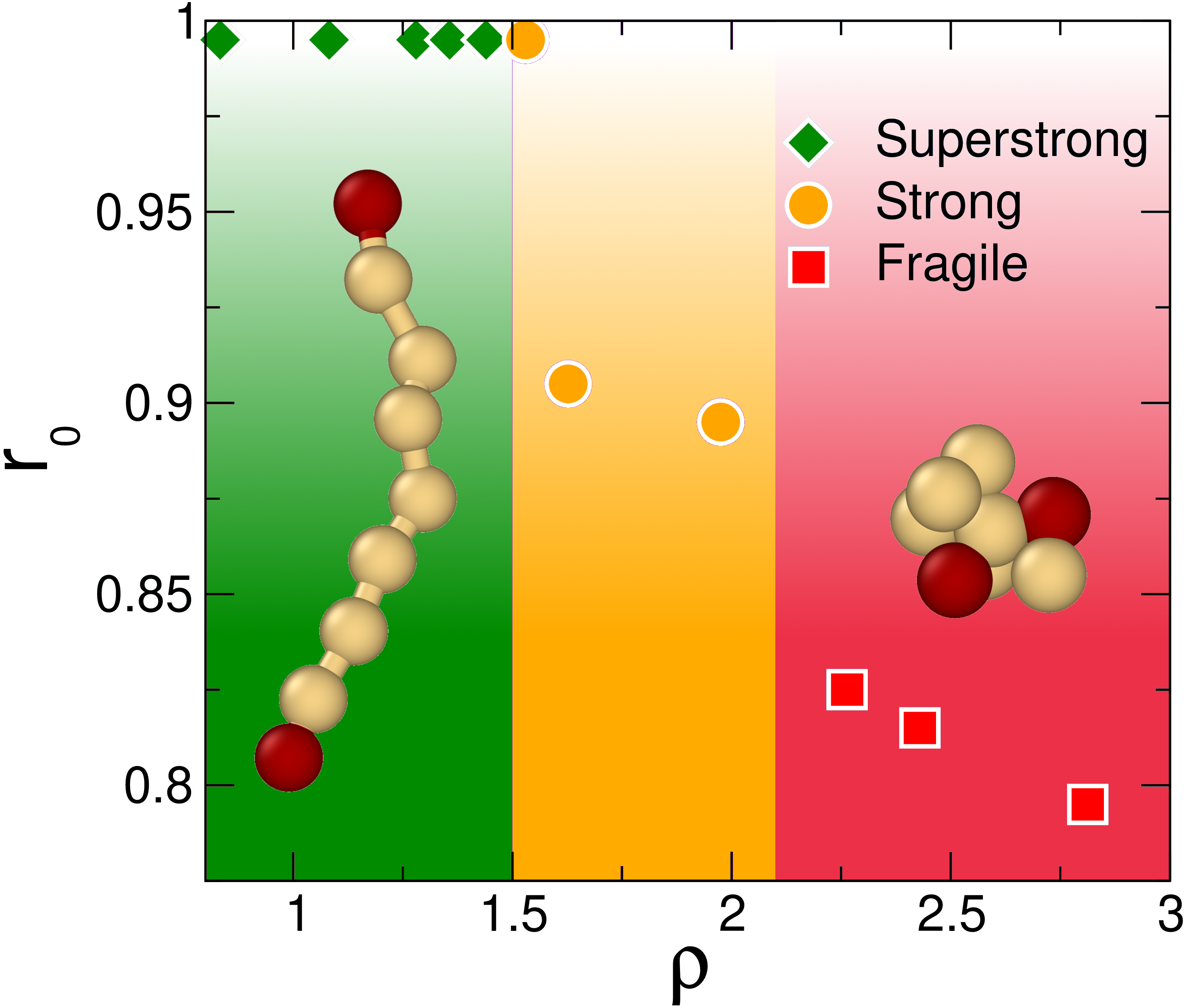}
\caption{Distance $r_0$ corresponding to the main peak of the radial distribution function $g(r)$ 
as a function of vitrimer bulk density. 
At low densities (green-shaded region), in which the system behaves as a dilute network of star polymers, 
the dominant length scale $r_0$ is set by the equilibrium length $L_0$ of the intra-star
harmonic bonds. As the density increases (orange-shaded region), excluded-volume interactions become more dominant,
resulting in a $r_0$ that is close to the repulsive WCA interaction range $\sigma=0.9 L_0$. 
At the highest densities studied (red-shaded region), segments are compressed to beyond the $\sigma$ interparticle distance,
resulting in $r_0 \approx 0.8 L_0$.}
\label{fig_peak}
\end{figure}

To gain a deeper structural understanding of the broad tunability of the vitrimer fragility
with bulk density, let us consider the real-space microstructure
as probed by the radial distribution function $g(r)$. Its main peak represents
the dominant nearest-neighbor interparticle distance $r_0$ that ultimately
governs the value of $S(k_0)$ and, consequently, the strength of the cage
effect and the sensitivity to temperature.
Figure~\ref{fig_peak} shows the values of $r_0$ as a function of vitrimer bulk
density at the respective glass transition temperatures $T_{g}$.
At high densities, in which case the vitrimers exhibit fragile behavior, $r_0$
corresponds to length scales smaller than $0.9$, where $0.9$ sets the range of our WCA
interaction potential.  That is, the dominant nearest-neighbor interactions
arise from steric repulsions among monomer segments, akin to the (fragile)
excluded-volume-dominated vitrification scenario in dense hard spheres~\cite{PhysRevLett.102.085703,Parisi2010}.
At low densities, however, $r_0$ approaches the equilibrium distance of the
harmonic-spring potential that connects the monomers within each star-polymer
arm.  In this limit, the intrinsic topology of the star-polymer system thus
plays a dominant role, and the structure resembles that of a strong network
material such as silica~\cite{horbach2001relaxation}. Since there are well defined
 ground states corresponding to fully bonded networks,
 the topology-dominated microstructure changes more
weakly than in the steric-repulsion-dominated case, and consequently 
the static structure factor saturates more quickly upon supercooling. Hence,
the low-density material behaves as a superstrong glass former.
Notice that the absence of a sharp transition for low-density low-valence networks
is consistent with the concept of ``equilibrium gels''~\cite{Sciortino2008} 
where the glass transition approaches $T=0$.
The intermediate-density regime falls in between these two extreme cases, combining
both topology-dominated (superstrong) and excluded-volume-dominated (fragile)
interactions to yield intermediate fragility indices. Lastly, this analysis
also explains why the value of the topology-freezing transition temperature
$T_v$ does not have a large influence on the fragility: reversible bond swaps
occur only at the ends of the star-polymer arms, while the vast majority of
monomer-monomer interactions---which ultimately control the dominant length
scale---are unaffected by such bond-swapping events.

\section{Schematic MCT analysis}
To unambiguously confirm whether the growth behavior of the main peak of the
static structure factor is indeed sufficient to classify the fragility of the
material, and to unite our theoretical results with the well-established
power-law divergence of microscopic MCT, we develop a schematic MCT model that
relies solely on the functional form of $S(k_0)$ upon supercooling.  Although
schematic MCT models inherently lack the full coupling of $k$-dependent density
modes, they are known to accurately reproduce many important features of fully
microscopic MCT, including the qualitative scaling behavior near the critical
point~\cite{Bengtzelius1984,Leutheusser1984}.  Following the full derivation described in the Supplementary Material,
we arrive at the following equation:
\begin{align}
\dot{\phi}(t)+\Omega^2(T)\phi(t)+\lambda(T)\int_0^t\phi^2(t-s)\dot{\phi}(s)ds=0.
\label{eq:spmct}
\end{align}
Here, $\phi(t)$ represents the normalized intermediate scattering function $F(k_0,t)/S(k_0)$, 
the dots denote derivatives with respect to time,
and the temperature-dependent functions $\Omega^2(T)$ and $\lambda(T)$ are given by
\begin{align}
\label{eq_ome}
\Omega^2(T)=& \frac{k_\mathrm{B}T}{m_0}\frac{k_0^2}{S(k_0;T)}
\\& \equiv C_1\frac{T}{S(k_0;T)} \nonumber
\end{align}
and
\begin{align}
\label{eq:lam}
\lambda(T)=& \frac{\sqrt{3}}{ 16\pi^2}\frac{k_\mathrm{B}T}{\rho m}\,k_0^3\,\left[S(k_0;T)-1\right]^2
\\ &\equiv C_2\, T\left[S(k_0;T)-1\right]^{2} \nonumber
\end{align}
where $m_0$ is the particle mass.
Based on the numerical vitrimer simulation data of Fig.~\ref{fig3}(c), we consider temperature-dependent
growth profiles of $S(k_0)$ of the form
\begin{equation}
    S (k_0;T)=1+\left(T^{-1}\right)^{1/\nu},
    \label{eq_s}
\end{equation}
where the exponent $\nu$ can be tuned to represent different growth behaviors
upon supercooling. Physically, this exponent thus captures the non-trivial role
of the vitrimer bulk density, effectively enabling a tuning from
steric-repulsion-dominated to topology-dominated interactions. For simplicity we solve Eqs.~\ref{eq:spmct}-\ref{eq:lam} for
$C_1=C_2=1$.

\begin{figure}
    \includegraphics[scale=0.4]{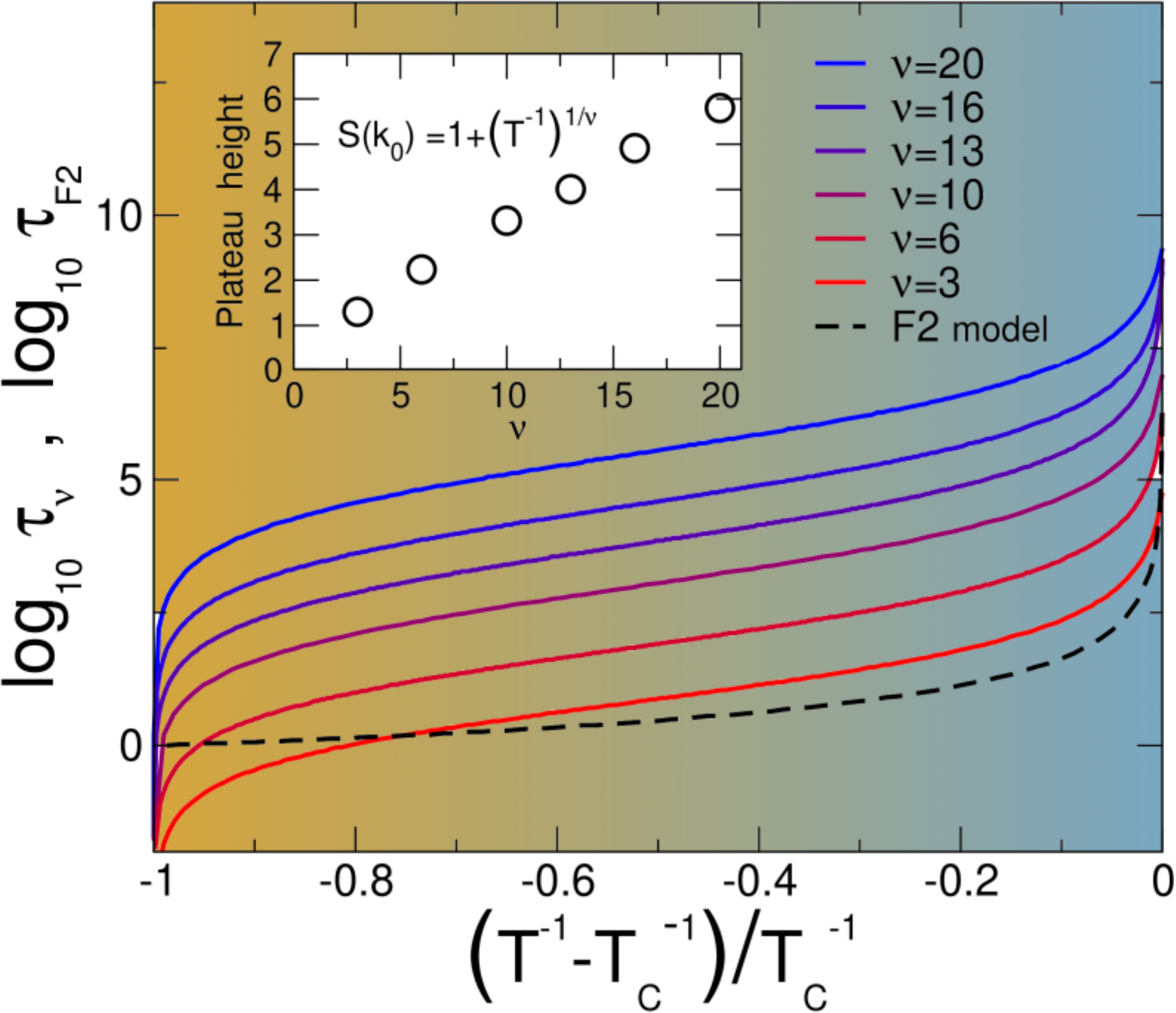}
    \caption{Angell plots of the relaxation time $\tau_{\nu}(T)$ as predicted from our schematic MCT model [Eqs.~\ref{eq:spmct}-\ref{eq:lam}]
	    , assuming
a peak growth of the static structure factor $S(k_0)$ as given by Eq.~(\ref{eq_s}) and $C_1=C_2=1$.
Different curves correspond to different values of the growth parameter $\nu$.
The dashed line indicates the results of the conventional $F_2$ reference model.
The glass transition temperature $T_c$ is defined as the point where the predicted relaxation time
strictly diverges.
The inset shows the plateau values at the inflection point, i.e.\ the $\tau_{\nu}$ values
at the temperature $T$ where the curvature of $\tau_{\nu}(T)$ changes sign.
Note that it grows exponentially with the parameter $\nu$.
}
    \label{fig_smct}
\end{figure}

Figure~\ref{fig_smct} shows the Angell plot of the calculated relaxation times
$\tau_{\nu}$ [defined via $\phi(\tau_{\nu})\equiv e^{-1}$] as a function of
inverse temperature. Note that the temperature scale in Fig.~\ref{fig_smct} is normalized with respect
to the corresponding schematic-MCT transition temperatures $T_{c}$, i.e.\
the temperatures at which $\tau_{\nu}$ diverges.  For comparison, we also plot
the predictions of the widely employed $F_2$
model~\cite{Bengtzelius1984,Leutheusser1984}, a simple schematic MCT model which assumes 
that $\Omega^2 \sim \textrm{constant}$ and $\lambda (T) \sim T^{-1}$. 
We use the fastest relaxation time of the $F_2$ model as our unit of time
to enable a fair comparison between the different models. 

There are two important conclusions to be drawn from the results in Fig.~\ref{fig_smct}. First,
an inflection point appears in $\tau_{\nu}(T)$ when using our $\nu$-dependent form of
the structure-factor peak. That is, there is an initial high-temperature regime
in which the relaxation time grows only weakly, but
as the temperature further decreases, a power-law divergence emerges
sufficiently close to the critical point $T_{c}$.
Note that such an
inflection point is absent in the conventional MCT $F_2$ model. Although our
microscopic MCT calculations for (super)strong vitrimers in Fig.~\ref{fig_4panels}(b) do not
exhibit any power-law behavior at the temperatures at which we can adequately
equilibrate our samples ($T>T_{g}$), our schematic MCT analysis reveals
that an MCT power-law divergence indeed exists, but only at temperatures
sufficiently close to the MCT critical point. We therefore unambiguously conclude that, 
for our vitrimer simulation model,  
the MCT critical point must lie below the
simulated glass transition temperature $T_{g}$. 

The second main observation is that, by increasing the value of $\nu$---which
physically corresponds to a weaker temperature variation of
$S(k_0)$---, the plateau value of $\tau_{\nu}$ at the inflection point is tuned to increasingly large values 
(see inset of Fig.~\ref{fig_smct}).  We now recall that the operational glass
transition temperature $T_g$ is defined as the temperature at which $\tau$
exceeds a certain threshold value (taken as $10^{4.5}\tau_0$ in simulation).
The tunability of $\nu$ thus provides an effective means to shift the
experimental window of observation, i.e.\ the temperature regime in which the
melt can still be equilibrated, such that only the logarithmic-like growth of
$\tau$ is manifestly visible.  In practice, this implies that an arbitrarily
slow growth of $S(k_0)$ can result in an arbitrarily large suppression of the cage effect and manifestly low fragility
indices $m$, as indeed we have found in Fig.~\ref{fig_4panels}(d).
Overall, this analysis confirms that the remarkably broad, density-controlled fragility range of our
vitrimer simulation model can be rationalized---within the context of
first-principles-based MCT---in terms of the broad variation of the growth behavior of $S(k_0)$.

\section{Discussion}
We have presented a combined numerical and theoretical study of the anomalous
fragility of glass-forming vitrimeric networks, a new class of recyclable polymer
networks endowed with reversible bond-swap functionality.  Our coarse-grained
numerical simulation model is composed of star polymers with intrachain
attractive harmonic bonds and interchain repulsive interactions; the end
segments of the polymer arms are functionalized such that they can undergo
reversible bond swaps. Our MD simulations reveal that, by changing the
bulk density $\rho$ by a factor of 2-3, the vitrimer fragility at $T_g$ can be tuned from fragile
(super-Arrhenius) to strong (Arrhenius) and even superstrong (sub-Arrhenius)
behavior. The fragility and corresponding density-dependent tunability are found to be independent of the topology-freezing
transition temperature $T_v$ at which all bond-swapping events cease, although
the value of $T_v$ does affect the absolute location of the glass-transition
line in the ($\rho, T$) phase diagram. 

Surprisingly, we can qualitatively reproduce the entire fragility range 
 using microscopic, fit-parameter-free MCT---a
first-principles theory that is conventionally assumed to be only applicable
to fragile glass formers. A detailed analysis allows us to trace the
microstructural origin of the vitrimer fragility to 
a single experimental observable: the main peak of the static
structure factor. We observe that the strongest vitrimers have the lowest variation
of $S(k)$ along the cooling protocol, which is particularly evident for the peak at wavenumber $k_0$: 
the weaker the growth of $S(k_0)$, the stronger the material.
We can subsequently rationalize this behavior in terms of a weakening of the cage effect and 
a change in the
vitrimers' dominant length scale, which switches from intra- to
interchain nearest-neighbor interactions as the density decreases. Finally,
based on these microscopic insights, we develop a simplified schematic MCT
model that takes into account only the growth behavior of $S(k_0)$. This model
not only recovers the qualitative features of our fully microscopic MCT
predictions, but it also reveals how our MCT-predicted fragility range can be
united with the theory's well-known fragile power-law divergence near the MCT
critical point. The latter lies below the operational glass transition
temperature of our simulation model, in the regime where the vitrimer melt has
already fallen out of equilibrium. Based on these results, we conclude that MCT
is in fact capable of predicting manifestly different fragilities, provided
that the MCT critical point lies sufficiently far below the experimental $T_g$.

Though not remedying the well-documented failures of standard MCT---which
ultimately stem from MCT's mean-field-like approximation for the memory 
kernel~\cite{gotze2008complex,Reichman2005,Liesbeth2018front}---our work 
uncovers a hitherto unknown regime in which
MCT serves as an accurate first-principles theory of glassy
dynamics. This provides a promising basis for further first-principles-based
studies of the glass transition, including the development of a Generalized MCT
that rigorously goes beyond the mean-field limit~\cite{SzamelPRL2003,JanssenPRE2014,Janssen2015a,Luo2019arxiv}. In future work, it will
be interesting to explore how our MCT analysis compares to other
non-fragile glass formers such as silica~\cite{horbach2001relaxation,horbach1999static,voigtmann2008dynamics,sciortino2001debye,PhysRevE.76.011507}, 
and to other glassy materials with
two or more competing molecular length scales. Furthermore, the anomalous superstrong fragility
may also be rationalized using other theoretical paradigms such as, e.g., the energy landscape \cite{debenedetti2001supercooled,Sussman2018},
which offers an intuitive picture for glassy dynamics in terms of energy-barrier-crossing events. 
 Finally, our findings, 
in particular the role of the density as a tunable fragility
parameter, and the anomalous growth behavior of the main peak of the static
structure factor, can be validated in rheological and scattering experiments of
vitrimer-like materials. Such a combined theoretical and experimental approach
holds exciting potential for the rational structural design of recyclable 
amorphous plastics with unprecedented dynamical and mechanical performance.

\matmethods{
\subsection*{Simulation details}
Every star polymer is composed of $n=25$ segments: one central segment and 8 arms of 4 segments each,
as depicted in Fig.~\ref{f_model}. Nearest-neighbor segments interact through a harmonic
potential with equilibrium length equal to the unit of length, and any other segment-segment
pair interacts through a repulsive WCA potential with cutoff radius $\sigma=0.9$.
For the end segments of every arm, we further distinguish between two different 
types, A and B (depicted as red and blue beads
in Fig.~\ref{f_model}). These segments are capable of forming A-B attractive bonds through a
generalized 20-10 Lennard-Jones potential of equilibrium length $2^{1/10}\sigma$ and bond energy $\epsilon$.
We study a mixture of $N_1=600$ stars with $7$ type-A ends and $1$ type-B end segment, 
and $N_2=300$ stars of the opposite composition (1 type-A and 7 type-B ends) for a total of $N=22500$ segments.
Since for all temperatures considered the bond energy $\epsilon$ is larger than the thermal energy $k_\mathrm{B}T$,
all simulations result in a fully bonded network of
$900$ star polymers with $2700$ A-B bonds and $1800$ unbounded ends of the majority type (A);
the latter can act as initiators for the swap reactions. 
Compared to Ref.~\cite{Ciarella2018}, the model was optimized in order to
avoid its rigidification at the topological transition $T_v$~\cite{denissen2016vitrimers} by changing the mixture ratios
to reduce the number of intra-star bonds. In this condition, the mechanism driving the dynamical arrest is
the standard glass transition, so that the system can relax and reach equilibrium even below $T_v$.

Vitrimeric bond-swap functionality is implemented by imposing both a 1-to-1 bond condition 
and a (A-B-A or B-A-B) swap mechanism through an additional three-body
potential $v_{ijk}^{(3)}$ ~\cite{sciortino2017three}:
\begin{equation}
    v_{ijk}^{(3)}=\lambda \epsilon \; v_{ij}^{bond}\left( r_{ij}\right) v_{ik}^{bond}\left( r_{ik}\right),
    \label{eq_3bp}
\end{equation}
where the indices $i,j,k$ run over all possible A-B-A or B-A-B triplets, and
$v_{ik}^{bond}$ represents the A-B (generalized Lennard-Jones) pair potential. 
The parameters are chosen to set swappable bonds of length
$r_{min}=2^{1/10}\sigma \approx 0.54$ and energy $\epsilon\gg k_\mathrm{B}T$. 
The
three-body parameter $\lambda$ effectively controls the swap rate through the energy
barrier $\Delta E_{swap}=\epsilon(\lambda-1)$.  Full details for the
potentials can be found in~\cite{Ciarella2018,Ciarella2019,sciortino2017three}.
In the low-temperature limit, swaps are required to reach true equilibrium
configurations~\cite{Oyarzun2018}, hence we keep $\Delta E_\text{swap}=0$ during equilibration.
The inclusion of swap events also accelerates the process of initial
self-assembly~\cite{Gnan2017}.

Our simulations are performed in a cubic box with periodic boundary conditions
within the $NVT$ ensemble. 
We start the network assembly and equilibration process at a relatively high temperature, $T=0.2$,
for several values of $\rho$.  
The bulk density $\rho$ is controlled by 
changing the size of the cubic box.
 Our production runs consist of $n_t=10^6$ to $10^8$ timesteps with a step size of
$dt=10^{-3}$, starting from the equilibrated samples.  From the configuration at
intermediate density $\rho=1.5 $ and $T=0.2$ we
measure the relaxation time $\tau_0=0.06$ that will be used as reference for the
dynamics.  We then proceed to decrease the temperature $T$ of every sample and
repeat the same protocol.  To confirm that our calculations are done at
equilibrium, we have verified that the temperature $T$, pressure $P$, and the structure factor 
$S(k)$ are constant along the production runs; to exclude any aging effects we have compared the 
self-intermediate scattering function at the beginning and at the end of each simulation run,
as show in Fig.\ S1. For every state point we have performed 
2 to 5 independent runs.
All simulations are performed using the HOOMD-blue
package~\cite{glaser2015strong}. 

We calculate the self- and collective intermediate scattering 
functions $F_s(k,t)$ and $F(k,t)$, as well as the static structure factors $S(k)$, 
from our MD production runs at different $\rho$ and $T$.  
Rather than using the full $25$x$25$ partial correlation matrices, 
we invoke the site-averaged approximation, which effectively replaces specific
monomer-monomer correlations with mean-field averaged ones. This choice of approximation is
motivated by scattering experiments in which it is not possible to
distinguish between the contributions of different monomers. Within the context of MCT,
the site-averaged approximation has also been
successfully applied to
polymers~\cite{chong2007structural,Baschnagel_2005,Aichele2004,chong2002mode,Sharpe1999,schweizer1997integral},
leading to a first-principles-based confirmation of the Rouse model~\cite{doi1988theory}.
Since we are interested only in qualitative results, we set the form factor $w(k)$~\cite{Wittmer_2007,Frey2015} to 1, thus 
ignoring the change in the conformation of our polymers. In Fig.\ S7 of the Supplementary Material, we test the effect
of the form factor by explicitly including it in our microscopic MCT calculations for a superstrong data set. 
The results confirm that the predicted fragility is unaffected by our site-averaged approximation.

\subsection*{Mode-Coupling Theory}
We solve the microscopic MCT equation-of-motion for the self- and collective intermediate scattering
functions $F_s(k,t)$ and $F(k,t)$ in the overdamped limit~\cite{Gotze1992,Reichman2005}. For the collective
dynamics the MCT equation reads
\begin{equation}
	\dot{F}(k,t) + \Omega^2(k)\,F(k,t) + \int M(k,t-\tau) \dot{F}(k,\tau) d\tau = 0
\label{eq:mct}
\end{equation} 
where $\Omega^2(k) = k^2k_\mathrm{B}T/[m_0S(k)]$ is the bare frequency term, $T$ the temperature, and $m_0$ the particle mass. 
The memory kernel in Eq.~\ref{eq:mct} is approximated in the standard MCT formalism as
\begin{equation}
\begin{split}
	M(k,t) = &(k_\mathrm{B}T/m_0)\rho/(16\pi^3) \int d\bm{q} |V_{\bm{k}-\bm{q},\bm{q}}|^2\;\cdot \\&  
                 \bigg[S(q) S(|k-q|) F(q,t) F(|\bm{k}-\bm{q}|,t)\bigg],
\end{split}
\end{equation}
where the vertex function $V_{\bm{k}-\bm{q},\bm{q}}=\bm{\hat{k}}\bm{q}c(q)+\bm{\hat{k}}(\bm{k}-\bm{q})c(|\bm{q}-\bm{k}|)$,
with $\bm{\hat{k}}=\bm{k}/|\bm{k}|$ and $c(q)$ the direct correlation function,
is determined by static quantities only. For $F_s(k,t)$ a similar MCT equation applies,
\begin{equation}
	\dot{F}_s(k,t) + \Omega_s^2(k)\,F_s(k,t) + \int M_s(k,t-\tau) \dot{F}_s(k,\tau) d\tau = 0,
\label{eq:mcts} 
\end{equation}
with $\Omega_s^2(k) = k^2 k_\mathrm{B}T/m_0$ and
\begin{equation}
	M_s(k,t) = (k_\mathrm{B}T/m_0)\rho/(16\pi^3) \int d\bm{q} (\bm{\hat{k}}\bm{q})^2 c(q)^2 F(q,t) F_s(|\bm{k}-\bm{q}|,t).
\end{equation}
The equations are solved under the standard boundary conditions $\dot{F}(k,0)=0$, $F(k,0)=S(k)$, $\dot{F}_s(k,0)=0$, and $F_s(k,0)=1$.
Time propagation is performed using Fuchs' algorithm ~\cite{Fuchs1991,Flenner2005}, 
starting with a step size of $dt=10^{-6}$ that doubles every 32 steps.
The $\bm{k}$-dependent integral in the memory kernel is solved by a double
Riemann sum on an equidistant grid of 100 wavenumbers, ranging from $k=0.2$ to $k$=40~\cite{Franosch1997}.
}

\showmatmethods{} 

\acknow{It is a pleasure to thank Francesco Sciortino, Barbara Capone, David Reichman, and Kees Storm for interesting discussions and Stefano Fine for technical support. 
L.M.C.J.\ acknowledges the Netherlands Organization for Scientific Research (NWO) for support 
through a START-UP grant.}

\showacknow{} 



\begin{thebibliography}{10}

\bibitem{montarnal2011silica}
Montarnal D, Capelot M, Tournilhac F, Leibler L (2011).
\newblock {\em Science} 334(6058):965--968.

\bibitem{capelot2012metal}
Capelot M, Montarnal D, Tournilhac FF, Leibler L (2012).
\newblock {\em J Am Chem Soc} 134(18):7664--7667.

\bibitem{denissen2016vitrimers}
Denissen W, Winne JM, {Du Prez} FE (2016).
\newblock {\em Chem. Sci.} 7(1):30--38.

\bibitem{Angell1995}
Angell CA (1995).
\newblock {\em Science} 267(5206):1924--1935.

\bibitem{Mattsson2009}
Mattsson J, et~al. (2009).
\newblock {\em Nature} 462(7269):83--86.

\bibitem{Sussman2018}
Sussman DM, Paoluzzi M, Marchetti MC, Manning ML (2018).
\newblock {\em EPL (Europhysics Lett.} 121(3):36001.

\bibitem{SokolovJCP2016}
Dalle-Ferrier C, et~al. (2016).
\newblock {\em J. Chem. Phys.} 145(15).

\bibitem{debenedetti2001supercooled}
Debenedetti PG, Stillinger FH (2001).
\newblock {\em Nature} 410(6825):259.

\bibitem{Dyre2006}
Dyre JC (2006).
\newblock {\em Rev. Mod. Phys.} 78(3):953--972.

\bibitem{Cavagna2009}
Cavagna A (2009).
\newblock {\em Phys. Rep.} 476(4-6):51--124.

\bibitem{Tarjus2011}
Tarjus G (2011) in {\em Dyn. Heterog. Glas. Colloids, Granul. Media}, eds.{}
  {L. Berthier, G. Biroli, J.-P.Bouchaud, L. Cipelletti}, van Saarloos W.
\newblock (Oxford University Press), pp. 39--67.

\bibitem{Berthier2011}
Berthier L, Biroli G (2011).
\newblock {\em Rev. Mod. Phys.} 83(2):587--645.

\bibitem{binder2011glassy}
Binder K, Kob W (2011).
\newblock (World Scientific).

\bibitem{Charbonneau2017}
Charbonneau P, Kurchan J, Parisi G, Urbani P, Zamponi F (2017).
\newblock {\em Annu. Rev. Condens. Matter Phys.} 8(1):265--288.

\bibitem{saika2001fragile}
Saika-Voivod I, Poole PH, Sciortino F (2001).
\newblock {\em Nature} 412(6846):514.

\bibitem{Sastry1998}
Sastry S, Debendetti PG, Stillinger FH (1998).
\newblock {\em Nature} 393(554):554--557.

\bibitem{Krausser13762}
Krausser J, Samwer KH, Zaccone A (2015).
\newblock {\em Proc. Natl. Acad. Sci.} 112(45):13762--13767.

\bibitem{Ozawa_2016}
Ozawa M, Kim K, Miyazaki K (2016).
\newblock {\em J. Stat. Mech. Theory Exp.} 2016(7):74002.

\bibitem{GnanNature}
Gnan N, Zaccarelli E (2019).
\newblock {\em Nat. Phys.} 15(7):683--688.

\bibitem{Asai2018}
Asai M, Cacciuto A, Kumar SK (2018).
\newblock {\em ACS Cent. Sci.} 4(9):1179--1184.

\bibitem{Yan6307}
Yan L, D{\"{u}}ring G, Wyart M (2013).
\newblock {\em Proc. Natl. Acad. Sci.} 110(16):6307--6312.

\bibitem{VanDerScheer2017}
{Van Der Scheer} P, {Van De Laar} T, {Van Der Gucht} J, Vlassopoulos D, Sprakel
  J (2017).
\newblock {\em ACS Nano} 11(7):6755--6763.

\bibitem{Philippe2017}
Philippe AM, et~al. (2018).
\newblock {\em Phys. Rev. E} 97(4):1--5.

\bibitem{doi:10.1063/1.4905472}
Williams I, Oğuz EC, Bartlett P, L{\"{o}}wen H, {Patrick Royall} C (2015).
\newblock {\em J. Chem. Phys.} 142(2):24505.

\bibitem{chong2007structural}
Chong SH, Aichele M, Meyer H, Fuchs M, Baschnagel J (2007).
\newblock {\em Phys. Rev. E} 76(5):51806.

\bibitem{Aichele2004}
Aichele M, Chong SH, Baschnagel J, Fuchs M (2004).
\newblock {\em Phys. Rev. E} 69(6):14.

\bibitem{Scopigno2003}
Scopigno T, Ruocco G, Sette F, Monaco G (2003).
\newblock {\em Science} 302(5646):849--852.

\bibitem{Royall2015}
Royall CP, Williams SR (2015).
\newblock {\em Phys. Rep.} 560:1--75.

\bibitem{Xia1999}
Xia X, Wolynes PG (2000).
\newblock {\em Proc. Natl. Acad. Sci.} 97(7):2990--2994.

\bibitem{Tarjus2005a}
Tarjus G, Kivelson SA, Nussinov Z, Viot P (2005).
\newblock {\em J. Phys. Condens. Matter} 17(50):R1143--R1182.

\bibitem{Sausset2008}
Sausset F, Tarjus G, Viot P (2008).
\newblock {\em Phys. Rev. Lett.} 101(15):155701.

\bibitem{gotze2008complex}
G{\"{o}}tze W (2008).
\newblock (OUP Oxford) Vol.{} 143.

\bibitem{Gotze1992}
Gotze W, Sjogren L (1992).
\newblock {\em Rep Prog Phys} 55(3):241--376.

\bibitem{Kob2002}
Kob W (2002) in {\em Slow Relaxations nonequilibrium Dyn. Condens. matter. Les
  Houches-Ecole d'Ete Phys. Theor.}, eds.{} {J. Barrat, M. Feigelman, J.
  Kurchan}, Dalibard J.
\newblock (Springer Berlin Heidelberg).

\bibitem{Gotze1999}
G{\"{o}}tze W (1999).
\newblock {\em J. Phys. Condens. Matter} 11(10A):A1----A45.

\bibitem{Mauro2014}
Mauro NA, Blodgett M, Johnson ML, Vogt AJ, Kelton KF (2014).
\newblock {\em Nat. Commun.} 5:1--7.

\bibitem{Ciarella2018}
Ciarella S, Sciortino F, Ellenbroek WG (2018).
\newblock {\em Phys. Rev. Lett.} 121(5):1--11.

\bibitem{Likos2006}
Likos CN (2006).
\newblock {\em Soft Matter} 2(6):478--498.

\bibitem{PhysRevLett.82.5289}
Watzlawek M, Likos CN, L{\"{o}}wen H (1999).
\newblock {\em Phys. Rev. Lett.} 82(26):5289--5292.

\bibitem{PhysRevLett.80.4450}
Likos CN, et~al. (1998).
\newblock {\em Phys. Rev. Lett.} 80(20):4450--4453.

\bibitem{Gu2017}
Gu Y, et~al. (2017).
\newblock {\em Proc. Natl. Acad. Sci.} 114(19):4875--4880.

\bibitem{sciortino2017three}
Sciortino F (2017).
\newblock {\em Eur. Phys. J. E} 40(1):3.

\bibitem{chong2002mode}
Chong SH, Fuchs M (2002).
\newblock {\em Phys. Rev. Lett.} 88(18):185702.

\bibitem{Sharpe1999}
Schweizer KS, Fuchs M, Szamel G, Guenza M, Tang H (1997).
\newblock {\em J. Environ. Monit.} 1(2):23N--25N.

\bibitem{schweizer1997integral}
Schweizer KS, Curro JG (1997).
\newblock (Advances in Chemical Physics) Vol.{}~98, pp. 1----142.

\bibitem{Liesbeth2018front}
Janssen LMC (2018).
\newblock {\em Front. Phys.} 6:97.

\bibitem{Charbonneau2011}
Charbonneau P, Ikeda A, Parisi G, Zamponi F (2011).
\newblock {\em Phys. Rev. Lett.} 107(18):1--5.

\bibitem{doi:10.1080/14786446408643668}
Maxwell JC (1864).
\newblock {\em London, Edinburgh, Dublin Philos. Mag. J. Sci.}
  27(182):294--299.

\bibitem{CALLADINE1978161}
Calladine CR (1978).
\newblock {\em Int. J. Solids Struct.} 14(2):161--172.

\bibitem{Phillips1985}
Phillips JC, Thorpe MF (1985).
\newblock {\em Solid State Commun.} 53(8):699--702.

\bibitem{rovigatti2018self}
Rovigatti L, Nava G, Bellini T, Sciortino F (2018).
\newblock {\em Macromolecules} 51(3):1232--1241.

\bibitem{Leutheusser1984}
Leutheusser E (1984).
\newblock {\em Phys. Rev. A} 29(5):2765--2773.

\bibitem{Landes2019}
Landes FP, Biroli G, Dauchot O, Liu AJ, Reichman DR (2019).
\newblock {\em arXiv:1906.01103}.

\bibitem{Coslovich2012}
Coslovich D (2012).
\newblock {\em J. Chem. Phys.} 138(12):1--9.

\bibitem{Frey2015}
Frey S, et~al. (2015).
\newblock {\em Eur. Phys. J. E} 38(2):11.

\bibitem{Baschnagel_2005}
Baschnagel J, Varnik F (2005).
\newblock {\em J. Phys. Condens. Matter} 17(32):R851----R953.

\bibitem{Charbonneau13939}
Charbonneau P, Ikeda A, Parisi G, Zamponi F (2012).
\newblock {\em Proc. Natl. Acad. Sci.} 109(35):13939--13943.

\bibitem{charbonneau2014hopping}
Charbonneau P, Jin Y, Parisi G, Zamponi F (2014).
\newblock {\em Proc. Natl. Acad. Sci.} 111(42):15025--15030.

\bibitem{roldan2017connectivity}
Rold{\'{a}}n-Vargas S, Rovigatti L, Sciortino F (2017).
\newblock {\em Soft Matter} 13(2):514--530.

\bibitem{PhysRevLett.102.085703}
Brambilla G, et~al. (2009).
\newblock {\em Phys. Rev. Lett.} 102(8):85703.

\bibitem{Parisi2010}
Parisi G, Zamponi F (2010).
\newblock {\em Rev. Mod. Phys.} 82(1):789--845.

\bibitem{horbach2001relaxation}
Horbach J, Kob W (2001).
\newblock {\em Phys. Rev. E} 64(4):41503.

\bibitem{Sciortino2008}
Sciortino F (2008).
\newblock {\em Eur. Phys. J. B} 64(3-4):505--509.

\bibitem{Bengtzelius1984}
Bengtzelius U, Gotze W, Sjolander A (1984).
\newblock {\em J. Phys. C Solid State Phys.} 17(33):5915--5934.

\bibitem{Reichman2005}
Reichman DR, Charbonneau P (2005).
\newblock {\em J. Stat. Mech. Theory Exp.} 2005(2):1--23.

\bibitem{SzamelPRL2003}
Szamel G (2003).
\newblock {\em Phys. Rev. Lett.} 90(22):228301.

\bibitem{JanssenPRE2014}
Janssen LMC, Mayer P, Reichman DR (2014).
\newblock {\em Phys. Rev. E} 90(5):52306.

\bibitem{Janssen2015a}
Janssen LMC, Reichman DR (2015).
\newblock {\em Phys. Rev. Lett.} 115(20):1--9.

\bibitem{Luo2019arxiv}
Luo C, Janssen LMC (2019).
\newblock {\em arXiv:1909.00428}.

\bibitem{horbach1999static}
Horbach J, Kob W (1999).
\newblock {\em Phys. Rev. B} 60(5):3169.

\bibitem{voigtmann2008dynamics}
Voigtmann T, Horbach J (2008).
\newblock {\em J. Phys. Condens. Matter} 20(24):244117.

\bibitem{sciortino2001debye}
Sciortino F, Kob W, Walter K, Kob W (2001).
\newblock {\em Phys. Rev. Lett.} 86(4):648.

\bibitem{PhysRevE.76.011507}
Berthier L (2007).
\newblock {\em Phys. Rev. E} 76(1):11507.

\bibitem{Ciarella2019}
Ciarella S, Ellenbroek WG (2019).
\newblock {\em Coatings 2019, Vol. 9, Page 114} 9(2):114.

\bibitem{Oyarzun2018}
Oyarz{\'{u}}n B, Mognetti BM (2018).
\newblock {\em J. Chem. Phys.} 148(11):114110.

\bibitem{Gnan2017}
Gnan N, Rovigatti L, Bergman M, Zaccarelli E (2017).
\newblock {\em Macromolecules} 50(21):8777--8786.

\bibitem{glaser2015strong}
Glaser J, et~al. (2015).
\newblock {\em Comput. Phys. Commun.} 192:97--107.

\bibitem{doi1988theory}
Doi M, Edwards SF (1988).
\newblock (Oxford University Press).

\bibitem{Wittmer_2007}
Wittmer JP, et~al. (2007).
\newblock {\em Europhys. Lett.} 77(5):56003.

\bibitem{Fuchs1991}
Fuchs M, G{\"{o}}tze W, Hofacker I, Latz A (1991).
\newblock {\em J. Phys. Condens. Matter} 3(1991):5047--5071.

\bibitem{Flenner2005}
Flenner E, Szamel G (2005).
\newblock {\em Phys. Rev. E} 72(3):1--15.

\bibitem{Franosch1997}
Franosch T, Fuchs M, G{\"{o}}tze W, Mayr MR, Singh AP (1997).
\newblock {\em Phys. Rev. E} 55(6):7153--7176.

\end{thebibliography}
\end{document}